\documentclass[conference]{IEEEtran}
\usepackage{hyperref}

\usepackage{setspace}
\IEEEoverridecommandlockouts
\ifCLASSINFOpdf
  \usepackage[pdftex]{graphicx}
\else
\fi

\usepackage{booktabs} 
\usepackage{amsmath,graphicx}
\usepackage{tabularx}
\usepackage{todonotes}
\usepackage{float}

\usepackage{amssymb,fixltx2e}
\usepackage{amsmath} 
\usepackage{rotating}
\usepackage{float}
\usepackage{pgfplots} 
\usepackage{xcolor}
\usepackage{subcaption}
\usepackage{float}
\usepackage{pgfplots} 
\usepackage{tikz}
\usepackage{tikzscale}
\usepackage{graphicx}
\usepackage{algorithm}
\usepackage[noend]{algpseudocode}
\usepackage{url}

\usepackage{amsthm}
 
\theoremstyle{definition}
\newtheorem{definition}{Definition}[section]

\begin{document}
\makeatletter

\def\@IEEEpubidpullup{8\baselineskip}

\makeatother

\title{CIMTDetect: A Community Infused Matrix-Tensor Coupled Factorization Based Method for Fake News Detection}

\author{
    \IEEEauthorblockN{Shashank Gupta\IEEEauthorrefmark{1}\thanks{Author is a Data Scientist at Flipkart, Bangalore}, Raghuveer Thirukovalluru\IEEEauthorrefmark{2}, Manjira Sinha\IEEEauthorrefmark{2}, Sandya Mannarswamy\IEEEauthorrefmark{2}}
    \IEEEauthorblockA{\IEEEauthorrefmark{1}IIIT-Hyderabad, India. 27392shashankgupta@gmail.com} \
    \IEEEauthorblockA{\IEEEauthorrefmark{2}Conduent Labs, Bangalore, India. \  {\{Raghuveer.Thirukovalluru, Manjira.Sinha, Sandya.Mannarswamy\}@conduent.com }}
    
}

%


\maketitle
\renewcommand\intextsep{6pt}
\renewcommand\dbltextfloatsep{6pt}
\renewcommand\dblfloatsep{6pt}
\setlength\tabcolsep{3pt}

\begin{abstract}
Detecting whether a news article is fake or genuine is a crucial task in today's digital world where it's easy to create and spread a misleading news article. This is especially true of news stories shared on social media since they don't undergo any stringent journalistic checking associated with main stream media. Given the inherent human tendency to share information with their social connections at a mouse-click, fake news articles masquerading as real ones, tend to spread widely and virally. The presence of echo chambers (people sharing same beliefs) in social networks, only adds to this problem of wide-spread existence of fake news on social media. In this paper, we tackle the problem of fake news detection from social media by exploiting the very presence of echo chambers that exist within the social network of users to obtain an efficient and informative latent representation of the news article. By modeling the echo-chambers as closely-connected communities within the social network, we represent a news article as a 3-mode tensor of the structure - $<$News, User, Community$>$ and propose a tensor factorization based method to encode the news article in a latent embedding space preserving the community structure.  We also propose  an extension of the above method, which jointly models the community and content information of the news article through a coupled matrix-tensor factorization framework. We empirically demonstrate the efficacy of our method for the task of Fake News Detection over two real-world datasets. Further, we validate the generalization of the resulting embeddings over two other auxiliary tasks, namely: \textbf{1)} News Cohort Analysis and \textbf{2)} Collaborative News Recommendation. Our proposed method outperforms appropriate baselines for both the tasks, establishing its generalization. 

\end{abstract}

%
\IEEEpeerreviewmaketitle

\section{Introduction}
Social media is increasingly becoming a popular platform for sharing and consumption of news. Due to its ever-rising popularity, more and more people tend to prefer it over the traditional news outlets. It is estimated that within the time period of 2012-16, there was a 26.5\% increase (from 49\% to 62\%) in the percentage of adult population consuming news from the social media\footnote{\url{https://goo.gl/rLFVP1}}. Amongst the younger generation (18-24 years old), it was observed that there is an increase of 16.7\% (from 24\% to 28\%) in the population who consume news from social media rather than from television\footnote{\url{http://www.bbc.com/news/uk-36528256}}. This popularity and ease of sharing has paved the way for digital misinformation propagation.

Owing to the phenomenon of information overload on social media, people are equally likely to share fake news articles as compared to factual news articles\footnote{\url{https://goo.gl/ZMXFUA}}. This phenomenon is reinforced by the confirmation bias - tendency to accept information/opinion confirming our own beliefs; and popularity bias - wherein an article is perceived of high quality and authentic, largely based on its popularity (number of shares for social media). These factors have led to a rise in fake news propagation on the social media.
Fake news has a negative impact on the balance of our society. It can influence public perception and create polarized views amongst the citizens. Fake news can be fabricated for political and financial gains. Recent times have witnessed a lot of controversy surrounding the influence of fake news on the U. S. presidential election \cite{allcott2017social}.

Echo-chambers are believed to have a great role in fake news propagation \cite{duggan2016political,del2015echo,menczer2016fake}.
Echo-chambers, which can be defined as a closely guarded community sharing same beliefs and opinions are an important concern over the social media. Some influential organizations and personalities have raised their concerns regarding it\footnote{\url{https://goo.gl/KDN8MN}}\footnote{\url{https://goo.gl/G2UE1t}}.  Echo-chambers are generally polarized and dense communities, making the misinformation spread almost instantaneous within the community. People often share it based on the headline alone, as long as it conforms with their pre-existing notions and beliefs (confirmation bias)\footnote{\url{https://goo.gl/DKdooU}}. 

In this work, we exploit the existence of echo chambers in social networks to obtain a latent representation of the news article, which can aid in fake news detection. We use a community enhanced tensor representation of the news article and propose a tensor factorization based method to generate its latent representation, which is then used for fake news classification. To the best of our knowledge, this is the first work which models the echo-chamber communities explicitly for the task of Fake News Detection. 

We posit a 3-mode tensor representation of the news - $<$News, User, Community$>$ to capture it's engagement with users and their communities on the social media followed by tensor factorization to generate a latent representation of news. Along with the community interaction of a news article, the content information also plays an important role in determining its veracity. It has been shown that the writing style of hyperpartisan (strong leaning) can be distinguished from mainstream news using certain linguistic features \cite{potthast2017stylometric}. We make use of this information and propose an enhancement of the above method, by combining the textual information of the news article with the community information in a coupled matrix-tensor factorization framework \cite{acar2011all}, which learns a compact latent representation encoding both the textual and community information. We posit that this enhanced information can be helpful in discriminating fake news from the genuine ones.
\subsection{Main Contributions}
Our core contributions can be summarized as follows: 
\begin{itemize}
\item We propose a first of its kind method which jointly models the echo-chambers with the news content information for the task of Fake News Detection.  
\item We propose a novel tensor factorization based method to learn a generic news embedding which is ultimately used in identifying fake news.
\item We also propose an extended Coupled Matrix-Tensor Factorization (CMTF) based method to jointly model the textual and community information of news article in a latent vector. We further demonstrate it's efficacy over content based and community based methods. 
\item We conduct extensive experiments over two real-world fake news datasets and demonstrate the superiority of our proposed method over established baselines. 
\item We show that the embeddings generated from our method are generic and hence can be used in other downstream tasks. To demonstrate it's generality, we design two auxiliary tasks: \textbf{1)} News Cohort Analysis and \textbf{2)} Collaborative News Recommendation. Our proposed method outperforms baseline methods over the two tasks described.
\end{itemize}

The rest of the paper is organized as follows: Section\ref{rel-work} presents the related works in fake news detection from social media. We introduce the necessary mathematical background in section \ref{foundation}. In section \ref{approach}, our proposed approach is described in detail. Section \ref{experiments} discusses the experimental details of our work followed by an analysis and conclusion in section \ref{results} and  section \ref{conclusion} respectively.

\section{Related Work}\label{rel-work} 

This section gives a brief summary of recent research in fake news detection methods. 
The existing literature on fake news detection models can be classified as follows:

\textbf{News Content Models:} Broadly, all the works in this category use content of the article to identify any fake news. Most news content models use the writing style and structure to determine the authenticity of news while some systems make use of an external database to verify the claims made in the article \cite{shu2017fake}. Rubin et al. use rhetorical structure theory to identify fake news based upon their coherence and struture. Yimin et al. \cite{chen2015misleading} employ a linguistic techniques to identify exaggerated, sensationalized content which leads to false news. Context free grammer rules were used to identify deception in Feng et al.\cite{feng2012syntactic} .Hassan et al. \cite{hassan2015detecting} uses content based features to identify check-worthy claims and verify them. Wu et al. \cite{wu2014toward} models claims as parametrized queries using structured data for verification with a database. 

\textbf{Social Context Models:} Social context based models make use of news' network graph - the users sharing the news and the propogation of the news on the social media platform to identify deceptive news. 
 Tacchini et al. \cite{tacchini2017some} used social context in the form of a bipartite network created from user ‘likes’ to identify hoax facebook posts. Jin et al. \cite{jin2016news} creates a credibility propagation network with tweet viewpoints and evolves it to verify news. Gupta et al. \cite{gupta2012evaluating} proposed an algorithm similar to pagerank on a multitype network of users, tweets and events to identify event credibility.
 
\textbf{News Content and Social Context Combined:} Most recent research combines traditional content based features other ancillary features to improve performance of the system. Ruchansky et al. \cite{ruchansky2017csi} which combines the news articles content features based RNN with social media engagement and user’s social context features network to decide the veracity of news article. Wu et al. \cite{wu2018tracing} use a LSTM-RNN network over social media propagation pathways of news to classify fake news. Jin et al. \cite{jin2017multimodal} proposed a deep architecture with an LSTM combining the text and social features and further  attention RNN to incorporate visual features into it, finally identifying fake news.

The presented research uses both - News Content information and Social Context information and proposes a novel methodology (CIMTDetect) to integrate the echo-chamber information and generate news article embeddings to identify fake news.

\section{Mathematical Terminology}\label{foundation}
In this section, we provide some background information for some of the techniques used later in the study. In particular, we will discuss tensors, their operations, followed by an introduction to tensor and coupled matrix-tensor factorization methods. 

\textbf{Tensors:} We assume that the readers are familiar with basic definitions of tensor and tensor products. Due to space constraints, we briefly define the relevant tensor/matrix operations needed for tensor factorization methods.

\begin{definition}{\textbf{Mode-n Matricization:}}
Matricization is defined as re-ordering of the elements of a tensor into a matrix. Formally, the mode-k matricization of a tensor $\mathit{X} \in \mathcal{R}^{I_1 * I_2 * ..... * I_N}$ is denoted by $\mathbf{X}_{(k)} \in \mathcal{R}^{I_n * \prod_{q \neq n}^{N} I_q}$, which arranges the mode-k fibers to be the columns of the resulting matrix.  
\end{definition}

\begin{definition}{\textbf{Mode-n Product:}}
The mode-n product of a tensor $\mathit{X} \in \mathcal{R}^{I_1 * I_2 * ..... * I_N}$ with a matrix $\mathbf{U} \in \mathcal{R}^{J * I_N}$  is denoted by $\mathbf{X} \times_k \mathbf{U}$, with $(\mathbf{X} \times_k \mathbf{U})_{i_1,....,i_{n-1},j,n+1,N} = \sum_{i_n = 1}^{I_n} x_{i_1,....,i_N} * u_{j, i_n} $. The dimensions of the new tensor are $I_1 * I_2 * ..... * I_{n-1} * J * I_{n+1} *.....*I_N$.
\end{definition}

\begin{definition}{\textbf{Khatri-Rao Product:}}
The Khatri-Rao product of two matrices $\mathbf{X} \in \mathcal{R}^{I * K}$ and $\mathbf{Y} \in \mathcal{R}^{J K}$ is defined as $\mathbf{X} \odot \mathbf{Y} = (x_1 \otimes y_1,...,x_K \otimes y_K )$, where $a_1,....,a_k$ and $b_1,....,b_K$ are the columns of the matrices $\mathbf{X}$ and $\mathbf{Y}$, respectively.

For a more detailed and comprehensive survey, interested readers can refer to the article by Kolda et. al \cite{kolda2009tensor}.

\end{definition}

\begin{table}
\centering
 \begin{tabular}{l | c  } 
 \hline
 \textbf{Symbol} & \textbf{Interpretation} \\
 \hline
 $x$ & scalar \\
 \hline
 $\mathbf{x}$ &  vector \\
 \hline 
 $\mathbf{X}$ &  matrix \\
 \hline 
 $\mathbf{\mathit{X}}$ &  tensor \\
 \hline $\mathbf{\mathit{X}_{i,j,k}}$ & Scalar-element at \{i,j,k\} index of tensor $\textbf{\textit{X}}$ \\
 \hline 
 $||\mathbf{X}||_F$ & Frobenious norm of matrix $\mathbf{X}$ \\
 \hline 
 $\mathbf{\mathit{X}} \odot \mathbf{\mathit{Y}}$ & Khatri-rao Product \\ 
\hline
 $\mathbf{\mathit{X}} \otimes \mathbf{\mathit{Y}}$ & Kronecker Product \\
 \hline 
\end{tabular}
\caption{Various symbols and their interpretation}\label{tab:resultsComb}
\vspace{-0.5cm}
\end{table}

\section{Proposed Method}\label{approach}
\subsection{Identifying Echo-Chambers}
Echo-chambers are closely connected community which shares common beliefs and opinions.  In typical lingua franca of social network analysis, echo-chambers are highly polarized groups(communities) in the network graph. Modularity score still being the most common metric for polarization in social networks \cite{gromping2014echo}, we use Girvan-Newman community detection algorithm \cite{newman2004fast} to identify communities representative of echo chambers in the network. It is a spectral method based on the modularity maximization principle. The modularity of a community is defined as the fraction of the edges that fall within the given community minus the expected such fraction if edges were distributed at random. The algorithm then proceeds by  starting with the full graph and then breaking it up into communities such that the modularity is maximized. The echo-chamber communities formed are used alongside other features in the following subsections.


\begin{figure*}[ht!]
\scriptsize
\centering
  \includegraphics[width=0.65\textwidth]{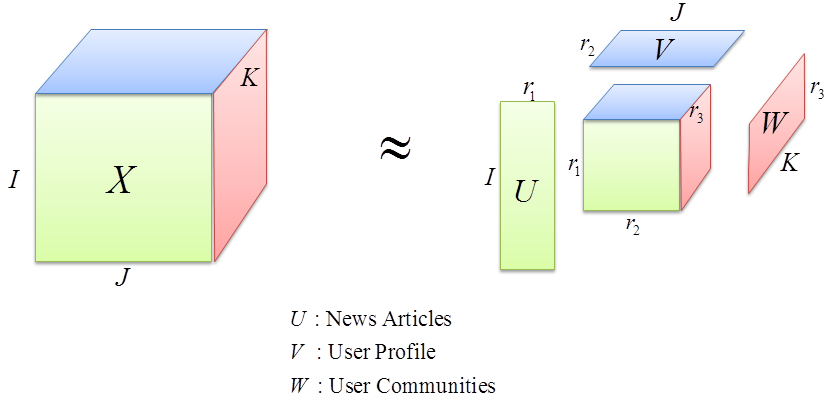}\label{mtl-dia}
  \caption{Graphical representation of the Tensor Factorization process with the three axes of the input tensor represented by: 1) News Articles, 2) User Profiles and 3) User Communities. Right side of the figure depicts the factors matrices which contains the lower dimensional embeddings of the respective entities along with the core tensor in the middle.}
  \label{TF}
\end{figure*}

\subsection{User-Community Matrix}
To encode the community structure, we construct a user-community matrix. The user-community matrix $C \in \mathcal{R}^{u \times c}$, where $u$ is the number of users in the social network graph and $c$ is the number of communities as identified by the community detection algorithm. $C$ is an indicator matrix where $C_{ij} = 1$, if user $i$ belongs to the community $j$ and $0$ otherwise. 

\subsection{News-Content Matrix}
To represent the news article's content, we use the n-gram count matrix. The n-gram count matrix $M \in \mathcal{R}^{n \times \left| V \right|}$, where $n$ is the total number of news articles and $\left| V \right|$ is the size of the vocabulary. Each entry $M_{ij} = c$  of the matrix depicts that the the frequency (n-gram count) of the $j$ n-gram in the news article $i$ is $c$.

\subsection{News-User Matrix}
To represent the news article's interaction with the social media, a news-user matrix is constructed. Each element of the news-user matrix $N \in \mathcal{R}^{n \times u}$, where $n$ is the number of news articles and $u$ is the number of users in the social graph present. Each element of the matrix $N_{ij}$ is a count variable indicating how many times user $j$ shared the news article $i$ on the social media.

\subsection{Tensor Formation}
\textbf{Infusing Community Information in News Representation:} Our main motivation behind this representation is to model the news article's interaction with the echo-chambers. Since in the real-world datasets, we don't have this information directly present, we encode this information through the interaction of the user who shared the news article with his community. We hypothesis that there is a distinctive pattern in how the news article is spread in the biased echo-chambers versus how its propagates otherwise. Going forward, we discuss how we represent the news article as a tensor.

Building on the definitions from the previous section, we give an overview on how the news is represented as a tensor. We represent a news article as a 3-mode tensor T - $<$news, user, community$>$, to capture the news article's interaction with the community (echo-chamber). Each entry of the tensor $t_{ijk}$ is computed as follows:

\begin{equation}
t_{ijk}  = N_{ij} * C_{jk}
\end{equation}

where $N_{ij}$ and $C_{jk}$ are defined previously. 

\subsection{Tensor Factorization}
Tensor factorization can be viewed as an higher-order extension of matrix factorization, where the objective is to decompose the higher-order tensor into low-rank tensors. As a result of the decomposition, the tensor can be expressed compactly as sum of lower-rank tensors. The resulting low-rank tensors captures complex interaction between the objects represented by the modes of the tensor. 

Tucker decomposition \cite{tucker1966some} can be seen as an higher-order extension of Principal Component Analysis (PCA). It decomposes the tensor into a core tensor along with matrices corresponding to each mode. Formally, a 3-way tensor $\mathit{X} \in \mathcal{R}^{I * J * K}$ is decomposed into factor matrices $\mathbf{U} \in \mathcal{R}^{I * r_1}$, $\mathbf{V} \in \mathcal{R}^{J * r_2}$ and $\mathbf{W} \in \mathcal{R}^{K * r_3}$ along with the core tensor $\mathit{G} \in \mathcal{R}^{r_1 * r_2 * r_3}$. The core tensor and the factor matrices are connected to the input tensor through the following relation:

\begin{figure}[ht!]
\centering
  \includegraphics[width=0.75\linewidth ]{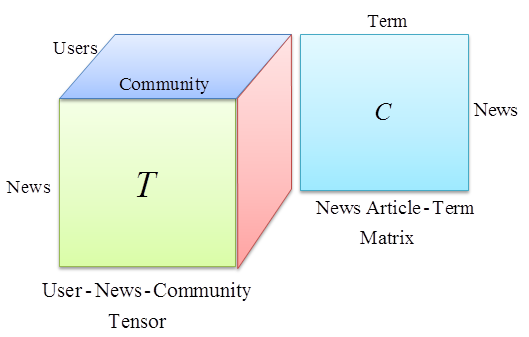}\label{CMTF}
  \caption{Cartoon representation of the coupled Tensor-Matrix representation of the news article. The matrix contains the term-frequency count over the vocabulary of the news article. Note that the first-mode of both the Tensor and the matrix, i.e. News Article is shared. }
  \label{fig:boat1}
  \vspace{-0.4cm}
\end{figure}

\begin{equation}
\mathit{X} \approx \mathit{G} \times_1 \mathit{U} \times_2 \mathit{V} \times_3 \mathit{W} 
 \end{equation}
 \begin{equation}
 \hspace{16mm}=\sum_{i=1}^{I}\sum_{j=1}^{J}\sum_{k=1}^{K} g_{i,j,k} u_i \circ v_j \circ w_k
\end{equation}
The following is the shorthand representation of the above equations:
\begin{equation}
\mathit{X} \approx [[\mathit{G};\mathbf{U},\mathbf{V},\mathbf{W}]]
\end{equation}

To find the factors mathematically, Higher-Order SVD (Ho-SVD) is used \cite{de2000multilinear}. Mathematically, HO-SVD attempts to solve the following optimization objective:

\begin{equation*}
\begin{aligned}
& \underset{G,U,V,W}{\text{minimize}}
& & || \mathit{X} - [[\mathit{G};\mathbf{U},\mathbf{V},\mathbf{W}]] \hspace{2mm} ||_F \\
\end{aligned}
\end{equation*}
with the additional constraint of the factor matrices being column-wise orthogonal. Computationally, a variant of Alternating Least Squares (ALS) algorithm called as TUCKALS3 is employed to find the factors for a 3-way input tensor \cite{kroonenberg1980principal}. 

The process is depicted in the Figure \ref{TF}. For our case, the input tensor is the tensor $\mathbf{T}$ as described previously. As result of the decomposition, we get latent representation of each entity, which are later used as feature for the final classification along with some auxiliary tasks.

\subsection{Fusing Content with Echo-Chambers}
In the previous section we described our approach to model news article's social interaction through the tensor formulation. In addition to the social network interaction, news article's textual content is also a strong signal in its final categorization, for obvious reasons. Therefore, we intent to combine the content signal with the social signal by jointly factorizing the article's content matrix ($\mathbf{C}$) with the news-user-community tensor ($\mathit{T}$). Doing so, we seek to jointly analyze ($\mathbf{C}$) and ($\mathit{T}$), decomposing them into latent factors coupled in the news dimension. Specifically, the first mode of $\mathbf{C}$) and ($\mathit{T}$, which represents the news article, share the same column sub-space.

\subsection{Coupled Matrix-Tensor Factorization}
Given the news-user-community tensor $\mathit{T}$ of dimensions $I * J *K$ and the term-matrix $\mathbf{C}$ of dimensions $I * |V|$, where $|V|$ is the size of the vocabulary, their joint factorization can be modeled as depicted in Figure 2. We follow the approach described by Acar et. al \cite{acar2011all} for the coupled Matrix-Tensor Factorization (CMTF). In CMTF, the tensor is factorized using the PARAFAC model \cite{harshman1970foundations}, also known as CP decomposition and the matrix is decomposed using the standard matrix factorization. Briefly, the CP decomposition of a tensor could be expressed as $\mathit{X} \approx [z] \times_1 \mathit{U} \times_2 \mathit{V} \times_3 \mathit{W}$, where $[z]$ is a superdiagnol tensor, with all of its element except the diagnol set to 0. 

\textbf{CMTF:} Given a tensor $\mathit{T} \in \mathcal{R}^{I * J * K}$ and a matrix $\mathbf{M} \in \mathcal{R}^{I * C}$, the objective function for CMTF is defined as follows:

\begin{equation}
f(\mathbf{U}, \mathbf{V}, \mathbf{W}, \mathbf{B}) =\frac{1}{2} || \mathit{T} - [[\mathbf{U}, \mathbf{V}, \mathbf{W}]] \hspace{2mm} ||^2_F + \frac{1}{2} || \mathbf{M} - \mathbf{U} \mathbf{B}^T ||^2_F  
\end{equation}

\begin{figure}[ht!]
\vspace{-0.25cm}
\centering
  \includegraphics[width=0.7\linewidth ]{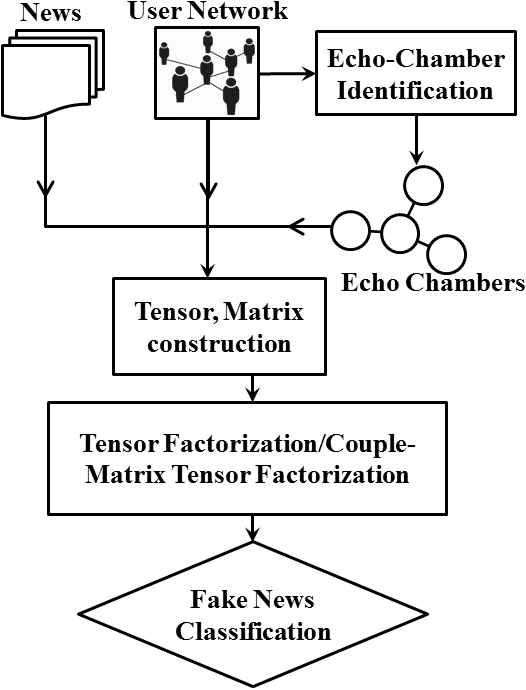}\label{CMTF}
  \caption{Overall system pipeline}
  \label{pipeline}
  \vspace{-0.2cm}
\end{figure}

The goal now is to find the factor matrices $\mathbf{B}, \mathbf{U}, \mathbf{V}, \mathbf{W}$ such that the objective function defined above is minimized. Any first-order optimization algorithm can be employed to find the solution. Rewriting the equation, 

\begin{equation}
f(\mathbf{U}, \mathbf{V}, \mathbf{W}, \mathbf{B}) = \frac{1}{2} f_1 + \frac{1}{2} f_2
\end{equation}

where we define $f_1 = || \mathit{T} - [[\mathbf{U}, \mathbf{V}, \mathbf{W}]] \hspace{2mm} ||^2_F$ and $f_2 = || \mathbf{M} - \mathbf{U} \mathbf{B}^T ||^2_F $. Next, we define the partial derivative of $f_1$ with respect to the factor matrices. Let $Z = [[\mathbf{U}, \mathbf{V}, \mathbf{W}]]$, then, 

\begin{equation}
\frac{\partial f_1}{\partial \mathbf{U}} = (Z_i - T_i) \mathbf{U}^{-i}
\end{equation}
where $\mathbf{U}^{-i} = \mathbf{U}^{I} \odot ... \mathbf{U}^{i+1} \odot \mathbf{U}^{i-1} \odot ..... \odot \mathbf{U}^1$. Similarly, the partial derivative of $f_1$ with respect to other factor matrices can be computed. 

The partial derivative of $f_2$ with respect to the factor matrix $\mathbf{U}$ can be computed as:
\begin{eqnarray}
\frac{\partial f_2}{\partial \mathbf{U}} = -MB + UB^TB \\
\frac{\partial f_2}{\partial \mathbf{U}} = -B^TU + BU^TU
\end{eqnarray}

Finally, combining the above results, the partial derivative of the initial objective function with respect to the factor matrix $\mathbf{U}$ can be computed as:
\begin{eqnarray}
\frac{\partial f}{\partial \mathbf{U}} = \frac{\partial f_1}{\partial \mathbf{U}} + \frac{\partial f_2}{\partial \mathbf{U}} 
\end{eqnarray}

Similar computations can be done for the remaining factor matrices. Readers interested in the full derivations of the above-mentioned partial derivatives can refer to the Acar et. al \cite{acar2011scalable}. By making use of the partial derivatives defined above, the optimization problem can be solved using any standard first-order optimization algorithms. The resulting factor matrices represent the lower-dimensional embedding of news articles, users and community, respectively, in our case. These lower-dimensional embeddings of news articles are then used as feature vector for our main task, Fake News Detection. This process is briefly summarized visually in Figure \ref{pipeline}.

\section{Experimental Details}\label{experiments}
\subsection{Datasets}
For evaluating the performance of our methods, we choose two real-world datasets with information such as news articles, social media users who shared the article along with the news article's content. The dataset is referred to as FakeNewsNet \cite{shu2017fake}\footnote{Dataset is available to download at \url{https://github.com/KaiDMML/FakeNewsNet}.}. Some statistics related to the dataset are presented in Table \ref{stats}. 

\begin{table}[h]
\centering
 \begin{tabular}{l | c  | c | c  } 
 \hline
  \textbf{Dataset} & \textbf{\#News Articles} & \textbf{\#Users} & \textbf{\#Fake-News Articles} \\
 \hline
 BuzzFeed & 182 & 15257 & 91 \\
 \hline
 PolitiFact & 240 & 23865 & 120  \\
 \hline 
\end{tabular}
\caption{Statistics of the Dataset}\label{stats}
\end{table}

The news articles are collected from two reliable news sources, BuzzFeed and PolitiFact. Both the datasets are balanced with respect to the output class, i.e. Fake News and Genuine News. 

\subsection{Experimental Settings}
To benchmark our method, we have considered a number of baselines, those can be broadly categorized into two types:
\textbf{Content-Based Baselines:}
These methods are based on the news article's textual content information only.
\begin{itemize}
\item \textbf{N-Gram + SVM:} This is a content based baseline with N-Gram count values as features followed by a SVM classifier.  

\item \textbf{NMF + SVM:} In this baseline, first the news article is represented by a N-Gram count matrix. Non-Negative Matrix Factorization (NMF) \cite{lee2001algorithms} is then used to embed the article in a latent semantic space. Classification is performed by using a SVM classifier. 

\item \textbf{SVD + SVM:} In this baseline, the news article is represented by a N-Gram count matrix, followed by a Singular Valued Decomposition \cite{golub1970singular} for its latent embedding. We use a SVM classifier for the final classification.

\item \textbf{RST:} RST \cite{rubin2015towards} uses  RST-VSM (Rhetorical Structure Theory and Vector Space Model) to embed the news article in a latent space. As part of RST, rhetorical structures, discourse constituent parts and their coherence relations are analyzed. Subsequently, a vector space model (VSM) is applied for embedding the news.

\item \textbf{LIWC:} LIWC \cite{pennebaker2015development} is a lexicon based method where the lexicon captures the psycholinguistic categories. The goal behind using such lexicon is to capture the deceptive features from the text. The assumption here is that deceptive features can discriminate intentionally written misleading article from the genuine one.  

\end{itemize}

\textbf{Content + Social Engagement Based Baselines:}
These approaches make use of both the content information of the news article along with its social media interactions by the users. 

\begin{itemize}

\item \textbf{Castillo:} The method proposed by Castillo et. al \cite{castillo2011information} uses new article's content and social engagement features. Several features are extracted from user's profile and his social network graph along with his credibility score. 

\item \textbf{RST + Castillo:} We combine features from RST and castillo, thereby considering both the news content and user's social engagements. 

\item \textbf{CITDetect:} This is our proposed method without the content information, i.e. using only the tensor representation of the news article. The embeddings generated from the tensor factorization algorithm is feed into a SVM classifier for the final classification. 

\item \textbf{CIMTDetect:} This is our proposed method with both the content information and the community-infused tensor information. The embeddings generated are feed into a SVM classifier for the final classification. 
\end{itemize}

\begin{table*}[ht!]
 \scriptsize
\centering
 \resizebox{\textwidth}{!}{%
 \begin{tabular}{|l | c | c | c |c | c | c| } 
 \hline
 \textbf{Method} & \multicolumn{3}{c|}{\textbf{BuzzFeed Dataset}} & \multicolumn{3}{c|}{\textbf{PolitiFact Dataset}} \\
 \hline
& \textbf{Precision} & \textbf{Recall} & \textbf{F1-score} & \textbf{Precision} & \textbf{Recall} & \textbf{F1-score} \\
  \hline
  N-gram + SVM &  0.523$ \pm$ 0.029 & 0.989 $\pm$ 0.044 & 0.684 $\pm$ 0.034 &  0.555 $\pm$ 0.043
 &		\textbf{0.933 $\pm$ 0.041}  & 0.696 $\pm$ 0.041 \\
 NMF + SVM &  0.586 $\pm$ 0.144 & 0.912 $\pm$ 0.150 &	0.711 $\pm$ 0.131 &  0.558 $\pm$ 0.032	&	0.925 $\pm$ 0.082	&	0.696 $\pm$ 0.039  \\
 SVD + SVM  & \textbf{0.843 $\pm$ 0.270} &	0.164 $\pm$ 0.094 & 0.268 $\pm$ 0.119
 &  \textbf{0.842 $\pm$ 0.191} &	0.408 $\pm$ 0.133	&	 0.549 $\pm$ 0.156 \\

RST  & 0.602 $\pm$ 0.66  &	0.561 $\pm$ 0.057 & 0.555 $\pm$ 0.057
 &  0.595 $\pm$ 0.032 & 0.533 $\pm$ 0.031	&	0.544 $\pm$ 0.042 \\

LIWC & 0.683 $\pm$ 0.065 & 0.628 $\pm$ 0.021 & 0.623 $\pm$ 0.066 & 0.621 $\pm$ 0.025 & 0.667 $\pm$ 0.091 & 0.615 $\pm$ 0.044 \\
\hline
 Castillo & 0.735 $\pm$ 0.08 & 0.783 $\pm$ 0.048 & 0.756 $\pm$ 0.051 &0.777 $\pm$ 0.051 & 0.791 $\pm$ 0.026 & 0.783 $\pm$ 0.015 \\
RST + Castillo & 0.795 $\pm$ 0.06 & 0.784 $\pm$ 0.074 & 0.789 $\pm$ 0.056 & 0.823 $\pm$ 0.040 & 0.792 $\pm$ 0.026 & 0.793 $\pm$ 0.032 \\  
  CITDetect (Proposed) & 0.657 $\pm$ 0.069 &	\textbf{1.000 $\pm$ 0.000} &	0.792 $\pm$ 0.051 & 0.679 $\pm$ 0.240
 &	0.975 $\pm$ 0.100 &	0.791 $\pm$ 0.111 \\
  CIMTDetect (Proposed) & 0.729 $\pm$ 0.116 &	0.923 $\pm$ 0.055 &	\textbf{0.813 $\pm$ 0.078} &  0.803 $\pm$ 0.132	&	 0.842 $\pm$ 0.133 &	\textbf{ 0.818 $\pm$ 0.069}\\
  
  \hline
\end{tabular}
 }
\caption{Accuracy Comparison for Various Methods (along with Std. Deviation)}\label{mainresults}
\end{table*}

\subsection{Data Preprocessing}
We have applied two types of pre-processing on the data.\\
\noindent \textbf{Text Pre-processing:} includes basic text pre-processing which involves lower-casing the text, stop-words and hyperlinks/url removal. For representing the text as term-matrix, we use the n-gram count representation where each element in the matrix is the n-gram count of the corresponding vocabulary term in the document. 

\noindent \textbf{Network Pre-processing:}
As part of the network pre-processing, after the community detection stage, we remove low-density communities, i.e. communities with less than some threshold number of members. 

\subsection{Hyper-parameter settings} 
We choose bi-grams for text representation. We limit the number of bi-grams to top 10k bigrams. We use the python library Sklearn for text-processing\footnote{\url{http://scikit-learn.org/}}. We make use of SVM with linear kernel as the final classifier. For community detection, the number of communities are decided empirically based on the cross-validation search. We use the SNAP tool for community detection\footnote{\url{https://github.com/snap-stanford/snap}}. We filter out communities with less than 5 members. The number of communities after the cross-validation search and filtering are 16 and 81 for the BuzzFeed dataset and PolitiFact dataset respectively. For both NMF and SVD factorization of the term-matrix, we set the dimensions of the lower-dimension embedding as 45. For tensor factorization we make of the python library Tensorly\footnote{\url{https://github.com/tensorly/tensorly}}. The dimensionality for the resulting news embedding is set to 45, for the social media users, it is set to 100, and for communities, it is set to 5 empirically. For CMTF, we make use of the MATLAB CMTF Toolbox\footnote{\url{http://www.models.life.ku.dk/~acare/CMTF_Toolbox}}. The dimensionality of the resulting embeddings is set to 45 for the BuzzFeed dataset and 75 for the PolitiFact dataset empirically.     

\section{Results and Discussion}\label{results}
 We use Precision, Recall and F1-score as classification evaluation metrics. We report results as the average of 5-Fold classification along with the standard deviation across the folds. The results are presented in Table \ref{mainresults}. The results from content-based methods are presented in the first half of the table, and the results from content + social based methods presented in the second half of the table. 

Amongst the content-based methods, NMF followed by a SVM classifier performs the best in terms of F1-score on both datasets. For the PolitiFact dataset, n-gram based representation performs similar to the NMF based method, in terms of F1-score. It is interesting to note that, n-gram based representation have the highest recall among the content-based method. CITDetect has a 100\% recall for the BuzzFeed dataset. Note that combining social and content features using existing techniques (Castillo, RST + Castillo) results in a significant improvement over the baselines using either of the features alone. \textbf{Finally, our proposed methods CIMTDetect performs best among all methods in terms of F1-score on both datasets.} 

\subsection{Evaluating the goodness of the embeddings}\label{analysis}




\begin{table}[t!]
\centering
 \begin{tabular}{|l | c | c | c |c|} 
 \hline
  & \multicolumn{2}{c|}{\textbf{BuzzFeed Dataset}} & \multicolumn{2}{c|}{\textbf{PolitiFact Dataset}} \\
 \hline
\textbf{Method} & \textbf{P@1} & \textbf{P@5} & \textbf{P@1} & \textbf{P@5} \\
  \hline
  UC-NMF & 0.16860  & 0.00157  & 0.16792     
 &		0.00150  \\
  \hline 
 CITDetect &  0.74791 & 0.01035 &	0.68719 &  0.00439	  \\
  \hline 
 CIMTDetect  & \textbf{0.93406} &	\textbf{0.01291} & \textbf{0.90626}
 &  \textbf{0.00469}	  \\

 \hline 
\end{tabular}
\caption{Results for evaluating Collaborative News Recommendation for users. }\label{reco}
\vspace{-0.45cm}
\end{table}

Our proposed methods CITDetect and CIMTDetect, give generic embeddings of the news articles and social media users as a result of factorization based on the content and social media/social network interaction. We have used the embeddings for our objective of Fake News Detection in the previous section. In this section, we aim to analyze the applicability and goodness of the embeddings for tasks other than Fake News Detection. Towards this, we design the following auxiliary tasks: \textbf{1)} Cohort analysis of the news articles and \textbf{2)} Collaborative News Recommendation.

\subsubsection{\textbf{News Cohort Analysis}} 
In this study, we test the hypothesis that a good representation (latent or explicit) should result in the formation of good clusters. We perform a qualitative analysis of the news representation learnt in terms of the quality of the clusters formed. In our case, we don't have an access to any human-labeled ground-truth clusters of the data, so we resort to unsupervised inter-cluster validation metrics.   

The different clustering metrics used here capture different \textit{goodness} measures of the clusters based on the inter-cluster and intra-cluster similarities. The Silhouette Index \cite{rousseeuw1987silhouettes} is calculated using the mean intra-cluster distance $(a)$ and the mean nearest-cluster distance $(b)$ for each sample. The Silhouette Index for a sample is now: $(b-a)/\max(a,b)$. The Calinski-Harabasz Index (CH-Index) \cite{calinski1974dendrite} is defined as ratio between the within-cluster dispersion and between-cluster dispersion. 

The results for the cluster evaluations are reported in the Table \ref{clust-poli}. K-Means is used as the algorithm of choice for clustering. We vary the number of cluster within the range of $[2-10]$ due to rather limited size of the datasets available. We report the clustering results as average of 10 different runs of K-Means algorithm with different random-seed selected for each run. 

We compare our proposed method with NMF (content feature based) as the baseline. It is clear from the tables that CIMTDetect outperforms the baseline in majority for the Silhouette Index and for all cases on the CH-Index. Thus, we have empirically demonstrated the effectiveness of our method on the first auxiliary task. Next, we describe experiments from the second auxiliary task. 

\subsubsection{\textbf{Collaborative News Recommendation}} 
For this study, we define the task by drawing inspiration from the field of recommendation systems, in particular the collaborative filtering technique, a popular choice of algorithm used in building recommendation systems. We formulate the task as, recommending news articles to a user based on his previous reading or sharing on social media pattern and reading/sharing pattern of similar users. We make use of the assumption that a good user-representation should club users of similar interest together - i.e. users who are indeed more representative of the current user. Specifically, for each user, we make use of his lower-dimensional embeddings generated from different methods and find similar users in that particular latent space. Next, we compute the frequency of the news articles shared by those users and recommend top-k most frequent news articles to that particular user. 

In the dataset, for each user, we have a list of news articles shared by him. We treat that list as gold-standard and compare the recommended news articles to the gold-standard. We use Precision@k as metric for evaluating the recommendation's quality. Briefly, for each recommended article in the ranked list (ranked according to frequency), we check if it's present in the goldset articles. In case of a hit, we increase the corresponding precision value. 

The results are present in Table \ref{reco}. We report Precision@1 and Precision@5 for both datasets. As part of the dataset, we have the social features for each user, which includes the user's basic profile information (age, gender, location etc.), content features of the tweets shared by the user etc. As baseline method for collaborative filtering, we apply Non-Negative Matrix Factorization over the user features (we call it UC-NMF) and use the resulting embedding as user's feature vector. It is clear from the table that CIMTDetect outperforms the baseline with a significant margin on both the evaluation metrics as-well as both the datasets.

\begin{table*}[ht!]
\centering
 \begin{tabular}{|l | c | c | c |c | c | c| } 
 \hline
  & \multicolumn{3}{c|}{\textbf{Silhouette Index}} & \multicolumn{3}{c|}{\textbf{CH-Index}} \\
 \hline
\textbf{\#Clusters} & \textbf{NMF} & \textbf{CITDetect} & \textbf{CIMTDetect} & \textbf{NMF} & \textbf{CITDetect} & \textbf{CIMTDetect} \\
  \hline
  2 & 0.558  & 0.388  & \textbf{0.673}     
 &		11.619  & 4.069 &  \textbf{15.152} \\
  \hline 
 4 &  0.42 & 0.453 &	\textbf{0.627} &  14.268	&	4.131	&	\textbf{18.787}  \\
  \hline 
 6  & \textbf{0.428} &	0.427 & 0.371
 &  13.017 &	4.337	& \textbf{ 21.351}	  \\

 \hline 
8  & 0.383  &	0.170 & \textbf{0.442}
 &   13.058	&	4.49 & \textbf{26.432} \\

 \hline 
10 & \textbf{0.300} & 0.029 & 0.241 & 19.575 & 4.71 & \textbf{30.911} \\
\hline
\end{tabular}
\caption*{}\label{clust-buzz}
\end{table*}

\begin{table*}[ht!]
\vspace{-0.4cm}
\centering
 \begin{tabular}{|l | c | c | c |c | c | c| } 
 \hline
  & \multicolumn{3}{c|}{\textbf{Silhouette Index}} & \multicolumn{3}{c|}{\textbf{CH-Index}} \\
 \hline
\textbf{\#Clusters} & \textbf{NMF} & \textbf{CITDetect} & \textbf{CIMTDetect} & \textbf{NMF} & \textbf{CITDetect} & \textbf{CIMTDetect} \\
  \hline
  2 & 0.632  & 0.617  & \textbf{0.787}     
 &		19.288  & 5.395 &  \textbf{26.399} \\
  \hline 
 4 &  \textbf{0.576} & 0.550 &	0.509 &  15.573	&	5.597	&	\textbf{28.604}  \\
  \hline 
 6  & 0.542 &	0.476 & \textbf{0.592}
 &  18.6 &	5.631	& \textbf{33.202}	  \\

 \hline 
8  & \textbf{0.509}  &	0.452 & 0.508
 &   18.184	&	6.054 & \textbf{35.698} \\
 \hline 
10 & 0.443 & 0.525 & \textbf{0.467} & 18.435 & 6.341 & \textbf{37.532} \\
\hline
\end{tabular}
\caption{Results for evaluating news article's clustering for the News Clustering task. Results are reported as average over 10 different runs of clustering algorithm. K-Means is used as the clustering algorithm. The first table presents the results for the BuzzFeed Dataset with the second table reporting results for the PolitiFact dataset.}\label{clust-poli}
\end{table*}

\subsection{Parameter Sensitivity}

\begin{figure}[htbp]
  \begin{minipage}[b]{0.45\linewidth}
    \centering
    \includegraphics[width=\linewidth]{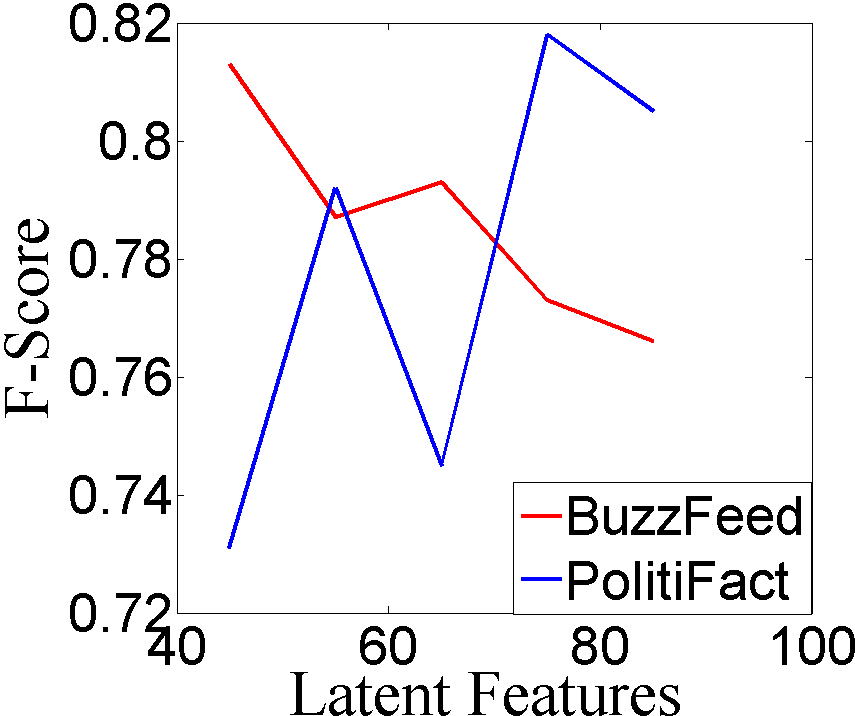}
    \label{fig:chapter001_dist_001}
    \vspace{-3mm}
  \end{minipage}
  \hspace{0.3cm}
  \begin{minipage}[b]{0.46\linewidth}
    \centering
    \includegraphics[width=\linewidth]{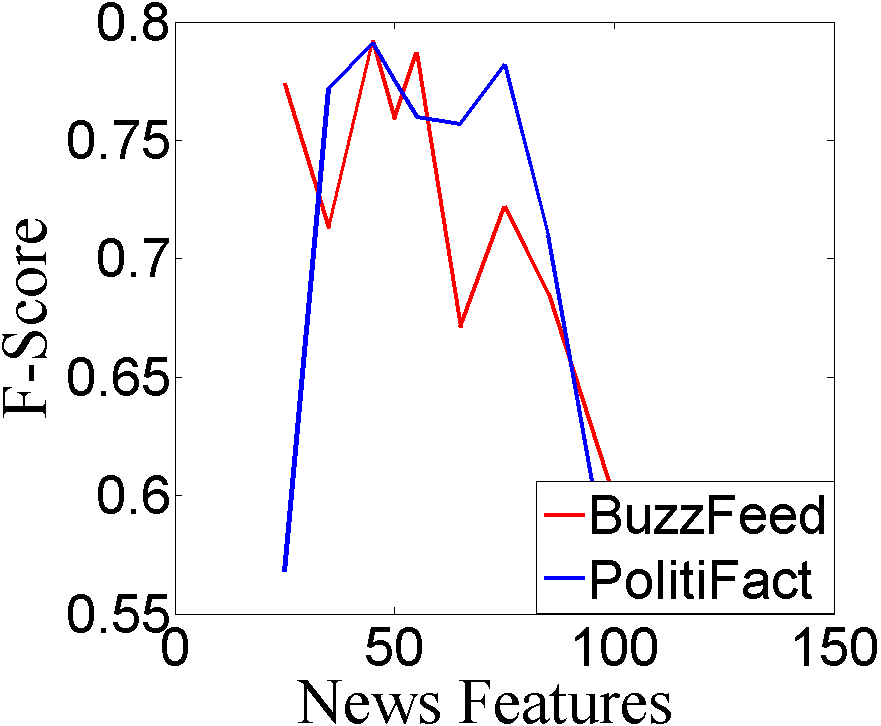}
    \label{fig:chapter001_reward_001}
     \vspace{-3mm}
  \end{minipage}
  \caption{Performance variation with news dimension for the CIMTDetect method (left figure) and CITDetect method (right figure)}
 
  \vspace{-2mm}
\end{figure}

We study the sensitivity of our proposed methods with respect to varying hyper-parameter. We analyze the variation in F1-score by varying the news embedding's dimension. Since our main entity of interest is the news article, we vary the latent dimension corresponding to the news article in both our methods, while keeping the other entities dimension constant. It is interesting to note that for CITDetect method, maximum F1-score is achieved for both datasets at the same dimensionality value of the news embedding. While in the CIMTDetect method, best F1-score is achieved at different dimensionality of the news embedding.  

\section{Conclusions}\label{conclusion}
In this paper, we present a systematic study analyzing the effects of modeling echo-chambers along with the content information for the task of Fake News Detection. We propose a Tensor Factorization based method (CITDetect) and its extension, a coupled matrix-tensor factorization based method (CIMTDetect) for the same. We find that modeling content information and echo-chamber (community) information jointly helps in improving the detection performance. We further propose two auxiliary tasks to verify the generalization of the our proposed methods and later demonstrate their effectiveness over baseline methods for the same. As part of future work, we plan to use Neural Network based methods for modeling echo-chambers for Fake News Detection.


\bibliographystyle{IEEEtran}
\bibliography{asonam_bib.bib}
\end{document}